\documentclass[sigconf]{acmart}
\usepackage{balance}       % to better equalize the last page
\usepackage{graphics}      % for EPS, load graphicx instead 
\usepackage{hyperref}
\usepackage{color}
\usepackage{booktabs}
\usepackage{textcomp}
\usepackage{subcaption}
\usepackage{enumerate}
\usepackage{xcolor}
\usepackage{lipsum}% http://ctan.org/pkg/lipsum
\usepackage{makecell}
\usepackage{multicol}
\usepackage{multirow}
\usepackage{array}
\usepackage{verbatimbox}
\usepackage{enumitem}
\usepackage{amsmath}
\usepackage{stfloats}
\usepackage{graphicx}
\usepackage{amsthm}
\usepackage{listings}
\usepackage{caption} 
\usepackage[export]{adjustbox}
\usepackage{xspace}
\usepackage{epsfig}
\usepackage[linesnumbered]{algorithm2e}
\usepackage{algpseudocode}
\usepackage{tabularx}
\usepackage{arydshln}
\usepackage[bottom]{footmisc}
\usepackage{tcolorbox}
\usepackage{stackengine}

\newcommand*{\eg}{\textit{e.g.},\xspace}
\newcommand*{\ie}{\textit{i.e.},\xspace}

\newcommand*{\etal}{\textit{et~al.}\xspace}

\newcolumntype{L}[1]{>{\raggedright\let\newline\\\arraybackslash\hspace{0pt}}m{#1}}
\newcolumntype{C}[1]{>{\centering\let\newline\\\arraybackslash\hspace{0pt}}m{#1}}
\newcolumntype{R}[1]{>{\raggedleft\let\newline\\\arraybackslash\hspace{0pt}}m{#1}}

\algnewcommand{\IfThenElse}[3]{% \IfThenElse{<if>}{<then>}{<else>}
  \State \algorithmicif\ #1\ \algorithmicthen\ #2\ \algorithmicelse\ #3}

\definecolor{darkblue}{HTML}{0C0893} 
\definecolor{brilliantlavender}{rgb}{0.6, 0.4, 0.8}
\definecolor{candypink}{rgb}{0.89, 0.44, 0.48}
\definecolor{lightpurple}{rgb}{0.8, 0.5, 0.98}
\definecolor{green}{rgb}{0.08, 0.47, 0.16}
\definecolor{violet}{rgb}{0.96, 0.5, 0.5}
\definecolor{asparagus}{rgb}{0.53, 0.66, 0.42}
\definecolor{darkpastelpurple}{rgb}{0.59, 0.44, 0.84}
\definecolor{mediumslateblue}{rgb}{0.48, 0.41, 0.93}

\definecolor{DarkGreen}{HTML}{5DAC81}
\copyrightyear{2023}
\acmYear{2023}
\setcopyright{rightsretained}
\acmConference[UbiComp/ISWC '23 Adjunct ]{Adjunct Proceedings of the 2023 ACM International Joint Conference on Pervasive and Ubiquitous Computing \& the 2023 ACM International Symposium on Wearable Computing}{October 8--12, 2023}{Cancun, Quintana Roo, Mexico}
\acmBooktitle{Adjunct Proceedings of the 2023 ACM International Joint Conference on Pervasive and Ubiquitous Computing \& the 2023 ACM International Symposium on Wearable Computing (UbiComp/ISWC '23 Adjunct ), October 8--12, 2023, Cancun, Quintana Roo, Mexico}\acmDOI{10.1145/3594739.3610677}
\acmISBN{979-8-4007-0200-6/23/10}

\begin{document}

%
% The "title" command has an optional parameter, allowing the author to define a "short title" to be used in page headers.
% \title{\projectname: Connecting Cough Detection and Lung Health Assessment with An End-to-End Deep Learning Model on Passively Sensed Audio}

\title{A Framework for Designing Fair Ubiquitous Computing Systems}

%
% The "author" command and its associated commands are used to define the authors and their affiliations.
% Of note is the shared affiliation of the first two authors, and the "authornote" and "authornotemark" commands
% used to denote shared contribution to the research.

\author{Han Zhang}
% \authornote{All authors are from the same organization.}
\email{micohan@cs.washington.edu}
\orcid{0000-0002-1377-1168}
\affiliation{%
  \institution{University of Washington}
  \city{Seattle}
  \country{USA}
 }

\author{Leijie Wang}
\email{leijiew@cs.washington.edu }
\orcid{0009-0007-2441-6018}
\affiliation{%
  \institution{University of Washington}
  \city{Seattle}
  \country{USA}
 }

\author{Yilun Sheng}
\email{ylsheng@cs.washington.edu}
\orcid{0009-0000-0263-2616}
\affiliation{%
  \institution{University of Washington}
  \city{Seattle}
  \country{USA}
 }

\author{Xuhai Xu}
\email{xuhaixu@uw.edu}
\orcid{0000-0001-5930-3899}
\affiliation{%
  \institution{University of Washington}
  \city{Seattle}
  \country{USA}
 }

\author{Jennifer Mankoff}
\email{jmankoff@cs.washington.edu}
\orcid{0000-0001-9235-5324}
\affiliation{%
  \institution{University of Washington}
  \city{Seattle}
  \country{USA}
 }

\author{Anind K. Dey}
\email{anind@uw.edu}
\orcid{0000-0002-3004-0770}
\affiliation{%
  \institution{University of Washington}
  \city{Seattle}
  \country{USA}
 }

%
% By default, the full list of authors will be used in the page headers. Often, this list is too long, and will overlap
% other information printed in the page headers. This command allows the author to define a more concise list
% of authors' names for this purpose.

%
% The abstract is a short summary of the work to be presented in the article.
\begin{abstract}
Over the past few decades, ubiquitous sensors and systems have been an integral part of humans' everyday life. They augment human capabilities and provide personalized experiences across diverse contexts such as healthcare, education, and transportation. However, the widespread adoption of ubiquitous computing has also brought forth concerns regarding fairness and equitable treatment. As these systems can make automated decisions that impact individuals, it is essential to ensure that they do not perpetuate biases or discriminate against specific groups. While fairness in ubiquitous computing has been an acknowledged concern since the 1990s, it remains understudied within the field. To bridge this gap, we propose a framework that incorporates fairness considerations into system design, including prioritizing stakeholder perspectives, inclusive data collection, fairness-aware algorithms, appropriate evaluation criteria, enhancing human engagement while addressing privacy concerns, and interactive improvement and regular monitoring. Our framework aims to guide the development of fair and unbiased ubiquitous computing systems, ensuring equal treatment and positive societal impact.
\end{abstract}

\begin{CCSXML}
<ccs2012>
   <concept>
       <concept_id>10003120.10003138</concept_id>
       <concept_desc>Human-centered computing~Ubiquitous and mobile computing</concept_desc>
       <concept_significance>500</concept_significance>
       </concept>
   <concept>
       <concept_id>10003120.10003138.10003139</concept_id>
       <concept_desc>Human-centered computing~Ubiquitous and mobile computing theory, concepts and paradigms</concept_desc>
       <concept_significance>500</concept_significance>
       </concept>
   <concept>
       <concept_id>10003120.10003138.10003142</concept_id>
       <concept_desc>Human-centered computing~Ubiquitous and mobile computing design and evaluation methods</concept_desc>
       <concept_significance>300</concept_significance>
       </concept>
 </ccs2012>
\end{CCSXML}

\ccsdesc[500]{Human-centered computing~Ubiquitous and mobile computing}
\ccsdesc[500]{Human-centered computing~Ubiquitous and mobile computing theory, concepts and paradigms}
\ccsdesc[300]{Human-centered computing~Ubiquitous and mobile computing design and evaluation methods}
%
% Keywords. The author(s) should pick words that accurately describe the work being
% presented. Separate the keywords with commas.
\keywords{Fairness; Ubiquitous Computing Systems; Framework}

\renewcommand{\shortauthors}{Zhang et al.}

%
% This command processes the author and affiliation and title information and builds
% the first part of the formatted document.
\maketitle

\section{Introduction}
\label{sec:introduction}

The proliferation of smart devices and the growth of the Internet of Things (IoT) have revolutionized the way we interact with technology and the world around us. These devices have made it increasingly feasible to capture human behavior passively and continuously, opening up new avenues for research and applications. Data of various types collected by these smart devices (\eg smartphones, smartwatches, and fitness trackers) have been used to design machine learning (ML) algorithms and systems that assess mental health \cite{saeb2015mobile, wang2014studentlife, xu2019leveraging, morshed2019prediction}, monitor cognitive load and work performance \cite{das2023algorithmic, nepal2022survey}, recognize human activities \cite{kwon2020imutube, peng2018aroma}, and track academic performance \cite{wang2015smartgpa, kim2021beneficial, bin2020measuring}. Ultimately, these systems play a pivotal role in aiding the decision-making process.

However, despite the potential for these algorithms and systems to improve people's lives, they may also introduce harm due to a lack of fairness considerations. The design of algorithmic decision-making systems can incorporate discriminatory biases, leading to unjust outcomes, especially for marginalized communities already facing societal inequalities~\cite{barocas2017fairness,pessach2022review}.
Therefore, it is crucial to ensure that the benefits of these systems are available to all, and that no group is unfairly disadvantaged. By addressing issues of fairness, we can deploy these technologies with better fairness, transparency, and accountability.

While fairness in ubiquitous computing has been acknowledged as a pressing concern since the 1990s \cite{weiser1993some}, research on fairness in the field of mobile, wearable, and ubiquitous computing has received comparatively less attention than in the field of machine learning. Recent work shows that only 5\% of papers published in the Proceedings of the ACM Interactive, Mobile, Wearable and Ubiquitous Technologies (IMWUT) journal from 2018 to 2022 included assessments of fairness, while most other studies focused on accuracy or error metrics \cite{yfantidou2023beyond}. The lack of attention to fairness in ubiquitous computing may be caused by several factors. First, unlike researchers in the machine learning fairness community, who typically use large-scale and publicly available datasets, ubiquitous computing researchers often collect small-scale and private author-collected datasets, which poses challenges to the design of common metrics and benchmarks for assessing fairness~\cite{xu2023globem}. Second, the sequential and dynamic nature of the data collected in ubiquitous computing (\eg steps and sleep patterns) makes it more difficult to identify biases from the surface compared to the more structured and static tabular format of data (\eg language corpus from social media) used by the machine learning fairness community \cite{yfantidou2023beyond}. Finally, the limited body of research on fairness in ubiquitous computing is also compounded by the lack of guidance for designing fair ubiquitous computing systems. Without such guidance, it is difficult to ensure that technologies designed for ubiquitous computing are being used in a fair and ethical manner.

\begin{figure*}
    \centering
    \caption{Overview of the framework for designing fair ubiquitous computing systems.}
    \includegraphics[scale=0.25]{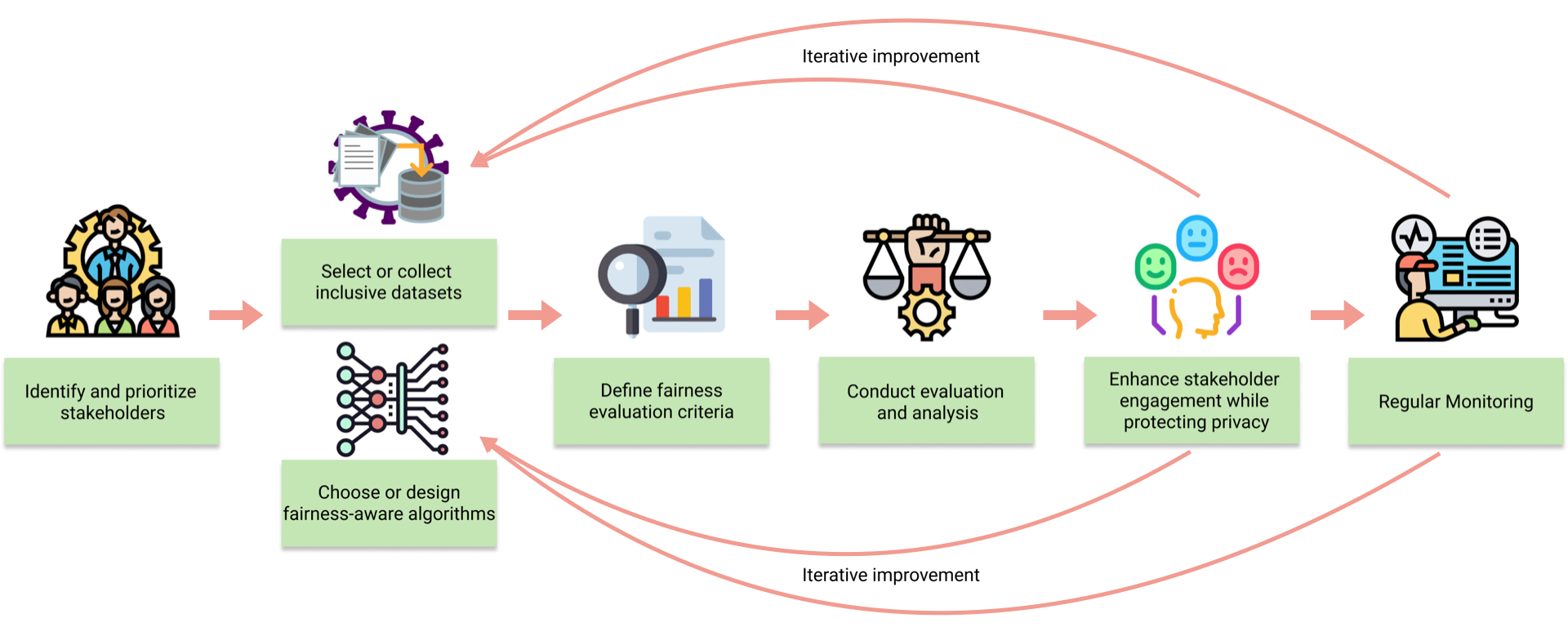}
    \label{fig:framework}
\end{figure*}

In this work, we propose a framework for designing fair systems in the context of ubiquitous computing. As illustrated in Figure \ref{fig:framework}, our framework starts by identifying relevant stakeholders in different contexts, determining who will use the algorithmic tool and who will be impacted by the algorithmic outcomes. Next, researchers gather appropriate and inclusive datasets and select or develop algorithms for evaluation. Evaluation criteria are then defined, incorporating fairness metrics and other performance measures, with a clear rationale for their selection (\eg why it is more reasonable to use the disparity of a traditional performance metric, such as false negative rate, as a fairness metric in a specific context). Algorithms are evaluated using the defined criteria and fairness metrics, and the results are analyzed within the context of stakeholder priorities. Transparency is ensured by communicating the findings and soliciting feedback from stakeholders. Finally, to adapt to the dynamic nature of ubiquitous computing, the systems are iteratively improved, and regularly monitored.

The contributions of this work are as follows. Firstly, we emphasize the significance of integrating fairness considerations into ubiquitous computing. Secondly, we present a framework that guides the design of fair ubiquitous computing systems. Lastly, we provide a detailed rationale for each component of the proposed framework. Our intention for this work is to serve as a valuable resource for future endeavors aiming to incorporate fairness into the design of ubiquitous computing systems.

\section{Background and Related Work}
\label{sec:background}

In this section, we first provide a broader review of prior work on fairness (Section \ref{related_work:algorithmic_fairness}). We then review the existing fairness literature in ubiquitous computing (Section \ref{related_work:ubicomp_lit}).

\subsection{A Broader Review of Existing Fairness Literature}\label{related_work:algorithmic_fairness}

As ML/AI is now being used in many decision-making systems, concerns have been raised about the fairness of these systems \cite{barocas2017fairness,zafar2017fairness}. In response to these concerns, Mehrabi \etal conducted a comprehensive systematic review of prior research, examining various sources of biases that can impact AI applications \cite{mehrabi2021survey}. Their study identified two key sources of unfairness in machine learning outcomes: biases originating from the data and biases arising from the algorithms themselves. To mitigate biases stemming from the data, researchers have proposed the adoption of inclusive benchmark datasets \cite{buolamwini2018gender}. These datasets aim to enhance the representation and diversity of the training data, thereby reducing the potential for biased and discriminatory outcomes in machine learning models.

Meanwhile, researchers have made substantial contributions to laying the foundation for understanding and mitigating algorithmic biases \cite{ hardt2016equality,dwork2012fairness,zafar2017fairness}. This work has resulted in the proposal of influential notions and frameworks aimed at promoting fairness in algorithmic decision-making systems.

One influential notion is \textbf{fairness through awareness}, proposed by Dwork \etal \cite{dwork2012fairness}. This notion emphasizes the consideration of \textbf{individual fairness}, which suggests treating similar individuals similarly. The authors highlighted the importance of ensuring ML models make consistent decisions and avoid discrimination based on protected attributes (such as race, gender, or age). Dwork \etal \cite{dwork2012fairness} also discussed \textbf{group fairness}, which focuses on fairness at the group level. They emphasized that the demographics of those receiving positive or negative classifications should align with the overall population. The concepts of individual fairness and group fairness contribute to the broader understanding and pursuit of fairness in machine learning and decision-making systems. 

To enforce algorithmic fairness, researchers have proposed various mathematical formulations and frameworks. Zafar \etal \cite{zafar2017fairness} explored the notion of \textbf{disparate mistreatment}, where ground truth is available for historical decisions used during the training phase. They provided a mathematical formulation for incorporating fairness criteria into the training process of machine learning models. This work allows practitioners to define and optimize fairness goals during modeling training, aiming to reduce bias in the resulting predictions. Another widely adopted framework, introduced by Hardt \etal \cite{hardt2016equality}, is \textbf{equality of opportunity}. This framework highlights the importance of equalizing the true positive rates across different demographic groups to ensure fairness. The authors provided theoretical analysis and practical algorithms for achieving equality of opportunity, and they demonstrated the effectiveness of their approach through empirical evaluations.

\subsection{Existing Ubiquitous Computing Fairness Literature}\label{related_work:ubicomp_lit}

While fairness research has made significant progress in addressing bias and discrimination in the machine learning fairness community, it is also crucial to consider these issues within the context of ubiquitous computing. Ubiquitous computing, characterized by the integration of computing power into everyday environments\cite{dey2001understanding,abowd1999towards,abowd2000charting}, presents unique challenges and opportunities for ensuring fairness. For example, one specific challenge arises from the dynamic nature of datasets collected in ubiquitous computing, which often makes it difficult to obtain an accurate understanding of the environment and capture every individual's needs. Consequently, biases and discrimination can be perpetuated and even amplified, leading to unfair outcomes for individuals. For instance, consider a health and fitness app that aims to provide personalized health recommendations based on user data collected from various wearable sensors. If the app's models are trained on outdated or inconsistent data due to irregular updates from users, it might provide inaccurate recommendations, potentially disadvantaging certain individuals.
% a smart campus environment that relies on location-based services might inadvertently favor groups who are closer to the campus by providing preferential access to resources based on their proximity, without considering the broader context of fairness. 

To address these concerns, researchers have recently begun to study fairness specifically in the field of ubiquitous computing. For instance, researchers conducted a systematic review of papers published in the IMWUT journal between 2018 and 2022 on algorithmic fairness \cite{yfantidou2023beyond}. They found that only 5\% of the published papers included fairness reports, indicating a need for more attention to fairness in the field of ubiquitous computing. Recently, researchers have started designing fairness-aware ubiquitous computing systems. One such work proposed a method that combines personalized federated learning with hierarchical clustering techniques to enhance the accuracy, robustness, and fairness of activity recognition systems \cite{li2023hierarchical}. By leveraging hierarchical clustering based on both activity similarity and user similarity, personalized models are created within smaller user groups, ensuring equitable treatment of individuals within the system. In another study, researchers developed a data-driven fairness-aware charging recommendation system specifically designed for large-scale electric taxi fleets \cite{wang2020faircharge}. The authors employed data-driven techniques and incorporated various self-defined fairness metrics (\eg reduction of traveling time and charging station occupation rates) to recommend charging stations to electric taxi drivers, aiming to ensure fairness in the allocation of resources.

\section{Towards Fair Ubiquitous Computing}
In this section, we provide a summary of the challenges and limitations that must be addressed to achieve fair ubiquitous computing (Section \ref{sec3:challenge}). This is followed by the objective of this work (Section \ref{sec3:goals}). 
% We emphasize the significance of addressing these challenges to ensure the development of fair ubiquitous computing systems 

\subsection{Challenges and Limitations towards Fair Ubiquitous Computing}\label{sec3:challenge}

Despite the emerging body of research that has started to investigate the fairness of systems in ubiquitous computing, as the example discussed in Section \ref{related_work:ubicomp_lit}, there are still several challenges and inherent limitations due to the unique nature of ubiquitous computing. Below, we identify six challenges and limitations and emphasize the need for addressing them.

\textbf{Limited consideration of sensitive attributes in specific contexts.} In contrast to the machine learning fairness community, which has explored a broad spectrum of sensitive attributes such as race, sexuality, disability, and nationality \cite{agarwal2018reductions, andrus2022demographic, calders2010three, kamiran2010discrimination, trewin2019considerations}, recent studies have highlighted a disparity in the focus of fairness work within the field of ubiquitous computing. Over the past five years, the research in ubiquitous computing has predominantly concentrated on gender and age attributes \cite{zhang2019pdmove, gong2022breathmentor, gullapalli2021opitrack}, with these attributes being mentioned in almost 90\% of the included papers \cite{yfantidou2023beyond}. On the other hand, many sensitive attributes, which are often associated with discrimination, have received limited attention in the ubiquitous computing literature \cite{lee2018detecting, manzini2019black,zhang2022impact}. Only a few of these papers discuss sensitive attributes through the lens of fairness.

For instance, one group of stakeholders that has been overlooked in the literature is individuals with disabilities \cite{stephanidis2019seven,emiliani2005universal}. Their unique needs require specific attention to ensure fair and inclusive ubiquitous computing experiences. 
Sexual orientation \cite{guz2021depression} is another important attribute that should be considered in fairness work for ubiquitous computing when modeling mental health. 

Given the wide implementation of ubiquitous computing across diverse contexts, the limited focus on different groups of stakeholders who may experience disproportionate effects from automated decision-making systems raises concerns and emphasizes the need for a broader consideration of these marginalized groups. To advance fairness and inclusively in ubiquitous computing, it is essential to expand the research scope and give careful consideration to identifying and prioritizing stakeholders in the specific context.

\textbf{Issue of Data Bias.} Similar to the machine learning fairness community, bias in datasets used for training models is a significant concern within ubiquitous computing community~\cite{mehrabi2021survey,yfantidou2023beyond}. These biases can stem from various factors, such as the data collection process, sampling techniques, or the inherent societal biases present in the data. Additionally, as ubiquitous computing systems extensively use data gathered from diverse sources, including sensors, mobile devices, and online platforms \cite{abowd1999towards,krumm2018ubiquitous}, these data sources often suffer from sparsity and uneven distribution, which can introduce biases and perpetuate existing inequities. 

The scarcity or imbalance of data can result in underrepresentation of certain groups, leading to biased algorithms and discriminatory outcomes. Therefore, it is crucial to address these challenges in data collection and ensure that data used in ubiquitous computing systems is representative, diverse, and less inherently biased.

\textbf{Limited Fairness-aware Algorithms in Ubiquitous Computing Systems.} Despite an emerging body of research that has started to investigate algorithmic fairness and mitigate bias in algorithms (\eg \cite{zafar2017fairness,hardt2016equality}), the development of fair algorithms for ubiquitous computing systems lags significantly behind. Based on the comprehensive review of fairness papers published in IMWUT between 2018 and 2022, it is evident that only a limited number of studies (three papers \cite{shahid2020leveraging,sheng2020weakly,zhang2021quantifying}) have explored the incorporation of fairness considerations into machine learning algorithms during the training process \cite{yfantidou2023beyond}. 

This can be attributed to the distinct challenges presented by ubiquitous computing, setting it apart from other domains. For instance, ubiquitous computing systems rely heavily on contextual information, including user location, preferences, and social interactions. Contextual factors can introduce additional complexities when determining fairness, as fairness considerations might vary based on the specific context and user characteristics. Moreover, ubiquitous computing systems often make real-time or near-real-time decisions based on continuous time-series data. Ensuring algorithmic fairness in these dynamic, time-sensitive scenarios adds a layer of complexity. 

% \textbf{Context Sensitivity.} Ubiquitous computing systems operate in dynamic and diverse environments \cite{abowd2000charting,weiser1999computer}, making it challenging to define and apply common fairness criteria that are contextually appropriate. The definition of fairness may vary based on the specific ubiquitous computing application, user groups, and cultural contexts. Balancing fairness requirements with context-specific considerations is a complex challenge in designing fair ubiquitous computing systems.

\textbf{Lack of Context-aware Evaluation Criteria.} In contrast to the machine learning fairness community, which commonly relies on standard fairness metrics such as demographic parity \cite{barocas2016big}, equalized odds \cite{zafar2017fairness}, and equal opportunity \cite{hardt2016equality}, the ubiquitous computing community often employs performance metrics such as accuracy and error rate \cite{yfantidou2023beyond}  without providing explicit justifications. Given the datasets used in the machine learning fairness community are static, while datasets used in ubiquitous computing are context-specific and sequential, it is crucial for ubiquitous computing researchers to carefully select appropriate fairness metrics that align with their specific contexts.

As an illustration, consider the case of modeling depressive behavior. In this instance, employing the disparity of false negative rates between prioritized stakeholders and others as a fairness metric proves more suitable than utilizing commonly employed metrics such as demographic parity and equalized odds commonly used in the machine learning fairness community. This selection is substantiated by existing research that demonstrates higher levels of mental health concerns among prioritized stakeholders, such as females with depressive symptoms \cite{givens2007ethnicity, mcfadden2016health, lucero2012prevalence}. By focusing on the disparity of false negative rates, which quantifies the variations in misclassification rates for individuals with depressive symptoms, researchers can more accurately capture the specific challenges and disparities faced by the prioritized stakeholders. In contrast, adopting standard machine learning fairness metrics, which primarily strive for equal treatment across groups by assuming comparable levels of depressive symptoms across genders in this example, may fail to adequately address the nuanced needs and disparities inherent to this particular context.

Additionally, another challenge in both the ubiquitous computing and the machine learning communities, \ie how to define a threshold that determines the point at which models are classified as unfair across different groups, should be addressed \cite{yfantidou2023beyond}. This absence of a well-defined threshold for determining fairness in ubiquitous computing models makes it challenging to assess whether a model is truly fair and not simply due to chance.

\textbf{Lack of Transparency and Explainability to stakeholders.} Fairness is closely intertwined with the principles of transparency and explainability. As AI-powered systems play an increasingly significant role in consequential decision-making, their explainability becomes essential for end-users to make informed and accountable decisions \cite{ehsan2021expanding,liao2023ai}. Over the past few years, significant advancements have been made in addressing this aspect within the machine learning fairness and human-computer interaction communities. 

For instance, machine learning researchers have deliberately opted for certain machine learning models, such as decision trees and linear models, which possess transparent structures and inherently provide interpretable explanations for their problems (\eg \cite{lakkaraju2016interpretable,caruana2015intelligible}). Additionally, there has been substantial progress in developing techniques for model interpretability and explainability such as LIME \cite{ribeiro2016should} and SHAP \cite{lundberg2017unified,lundberg2020local}. 

HCI researchers have been focusing on designing machine learning models' outputs in a transparent and understandable manner for end-users to bridge the gap between complex machine learning models and human comprehension (\eg \cite{abras2004user,zhu2018value}). Additionally, researchers have been investigating ways to incorporate user feedback and control mechanisms in machine learning models to enhance transparency and trust \cite{preece2015interaction}.  

In contrast to other research communities, the transparency and explainability of existing work in the field of ubiquitous computing pose unique challenges. One primary reason is that ubiquitous computing systems often operate on vast and diverse datasets comprising heterogeneous and time-series data from multiple sources. The processing and analysis of such intricate data necessitate the utilization of sophisticated algorithms and models, rendering it arduous to elucidate their underlying rationale in a clear and interpretable manner. Furthermore, these systems commonly handle sensitive user data, including personal health information \cite{li2020extraction}, location data \cite{xu2019leveraging}, or behavioral patterns \cite{xu2023globem}. Maintaining privacy and security is of utmost importance in such contexts, often entailing practices such as data anonymization, access restrictions, and the careful selection of interviewees. However, this pursuit of privacy introduces a trade-off between fairness and privacy \cite{chang2021privacy}, further exacerbating the complexity and challenges of transparency and explainability.

\textbf{Need for Regular Monitoring.} In contrast to other domains where data may be relatively static, ubiquitous computing systems require regular monitoring for fairness due to their dynamic and adaptive nature \cite{schmidt1999there,dey2001understanding}. These systems collect data dynamically in real-time, which introduces the need for ongoing monitoring to detect any biases that may emerge as the system adapts and learns from new information. Contextual factors such as location, time, and social surroundings play a significant role in ubiquitous computing systems, influencing their behavior. Regular fairness monitoring becomes essential to ensure fair treatment across different contexts and prevent biases from impacting users. Additionally, ubiquitous computing systems often make real-time decisions and interact directly with users. Monitoring helps identify any biases or discriminatory patterns in these decisions and interactions, enabling corrective actions to be taken promptly. Furthermore, the adaptability of ubiquitous computing systems and their potential to amplify biases emphasize the necessity of regular monitoring to ensure that biases are not perpetuated or amplified, safeguarding fairness and equitable outcomes for users.

\subsection{Objective of This Work}\label{sec3:goals}

Our work is motivated by the challenges and limitations pertaining to fairness in ubiquitous computing. Our objective of this work is to adapt and integrate the existing frameworks and concepts (reviewed in Section \ref{related_work:algorithmic_fairness}) into a specialized framework tailored for ubiquitous computing. Through this work, we aim to advance the development and deployment of ubiquitous computing systems that prioritize fairness and effectively cater to the diverse needs of stakeholders. By advocating for the integration of fairness considerations into system design, we seek to pave the way for future research towards fair ubiquitous computing.

\section{Framework for Designing Fair Ubiquitous Computing Systems}
\label{sec:framework}

In this section, we present an overview of our proposed framework (shown in Figure \ref{fig:framework}), designed specifically to address the challenges and limitations discussed in the preceding section (Section~\ref{sec4:framework}). Additionally, we delve into potential avenues for future research and development (Section \ref{sec4:future_work}).
% and provide a high-level overview of its components.

\subsection{Overview of Proposed Framework}\label{sec4:framework}
Our framework has six components in total, each representing one important stage in ubiquitous computing system implementation:

\begin{itemize}
    \item[(1)] \textbf{Identify and prioritize stakeholders}. Identify relevant stakeholders, such as those who will use the systems, as well as those who will be affected and biased by the algorithmic outcomes. For example, in the context of education, students with minoritized identities may face biased outcomes, making them important stakeholders. Transparency is crucial for both students and instructors who will use the systems. In contrast to the existing prevalent ubiquitous computing fairness literature, which often concentrates on a restricted range of sensitive attributes, our framework highlights the crucial importance of meticulously considering diverse sensitive attributes within varying contexts, \eg taking sexual minority status into account when modeling mental health.
    \item[(2a)] \textbf{Select/collect inclusive datasets}. Collect or select representative datasets that include prioritized stakeholders for evaluation, based on the specific context. In comparison to the prevailing ubiquitous computing fairness literature that frequently relies on unrepresentative datasets, our framework underscores the importance of gathering more inclusive datasets.
    One example is the recently published GLOBEM dataset~\cite{xu2022globem}, where researchers intentionally over-sampled diverse subpopulation groups (\eg gender, race, and immigration).
    \item[(2b)] \textbf{Choose/design fairness-aware algorithms}. Carefully select or design explainable fairness-aware algorithms that are relevant to the identified context and align with the goals of the evaluation. In contrast to the prevailing ubiquitous computing fairness literature, which often neglects the consideration of fairness during the algorithm design process, our framework emphasizes the need for developing fairness-aware algorithms. Note that, step 2a and step 2b can be interchanged.
    \item[(3)] \textbf{Define evaluation criteria}. Determine appropriate fairness metrics and thresholds to quantify differences and biases across stakeholder groups. Provide explicit justifications for the chosen metrics and thresholds. For example, in the evaluation of a depression detection algorithm, the disparity in false negative rates across different groups serves as a critical fairness metric. This metric reflects the algorithm's failure to identify depression in certain populations, making it essential to address. To ensure that observed disparities are not merely due to chance occurrences, a statistical test can be conducted. In contrast to the existing literature on fairness in ubiquitous computing, which often lacks justification for the selected fairness criteria, our framework emphasizes the necessity of designing context-aware evaluation criteria.
    \item[(4)] \textbf{Conduct evaluation and analysis}. Use the selected fairness metrics to evaluate algorithms on selected datasets, and analyze the results based on the predetermined evaluation criteria within the specific context. Additionally, thoroughly discuss the potential harm to stakeholders that may arise from the algorithmic decisions. In contrast to the existing literature on fairness in ubiquitous computing, which often neglects the discourse on the harm to stakeholders, our proposed framework seeks to bridge this gap.
    \item[(5)] \textbf{Enhance stakeholder engagement while protecting privacy}. Communicate evaluation findings, recommendations, and potential limitations to stakeholders, fostering transparency, accountability, and stakeholder involvement in algorithmic decision-making. Safeguard stakeholder privacy throughout the process. In comparison to the prevailing literature on fairness in ubiquitous computing, which often overlooks the involvement of humans in the system design, our framework emphasizes the importance of including human perspectives in the design process. Additionally, our framework recognizes the significance of striking a balance between fairness and privacy considerations.
    \item[(6)] \textbf{Iterative improvement and regular monitoring}. Refine algorithms to address potential biases and unfairness. Iterate on algorithmic design, data collection, and preprocessing to enhance fairness. Continuously monitor real-world performance and update algorithms and evaluation processes to align with evolving fairness standards and best practices. In comparison to the prevailing literature on fairness in ubiquitous computing, which often lacks this step, our framework emphasizes the importance of iterative improvement and regular monitoring. 
\end{itemize}

\subsection{Potential Directions for Future Work}\label{sec4:future_work}

In this section, we outline potential directions for future work and extensions of the proposed framework for designing fair ubiquitous computing systems. These avenues of research offer opportunities to advance the framework and enhance its applicability in real-world contexts. 

\subsubsection{Validation and Case Studies} To further validate and demonstrate the effectiveness of the proposed framework, future work can focus on conducting validation studies and case studies in real-world scenarios. For example, researchers can apply the framework to specific contexts and assess its practicality. By selecting representative use cases, researchers can demonstrate how the framework can be implemented and tailored to address fairness challenges in different scenarios. Case studies can involve evaluating or deploying fair algorithms in healthcare settings and educational environments. The findings from these case studies will provide valuable insights into the framework's feasibility, efficacy, and adaptability across diverse ubiquitous computing application domains. Moreover, to enrich the validation process, future work should include interviewing identified stakeholders to gather their feedback on the framework and the results of fairness testing in various scenarios. This qualitative feedback can provide additional context and perspectives, further refining and validating the proposed framework. By conducting these studies, it can also contribute to the identification of potential challenges and limitations of the framework. 

\subsubsection{Balancing Privacy and Fairness} As ubiquitous computing systems rely on collecting and analyzing vast amounts of personal data to make algorithmic decisions, addressing the tradeoff between privacy and fairness stands as a significant future direction within the context of the proposed framework for designing fair ubiquitous computing systems. One possible avenue for future research lies in the exploration of strategies and methodologies to reconcile the inherent tension between privacy and fairness. Researchers can develop privacy-preserving algorithms and techniques tailored specifically to ubiquitous computing environments and delve into privacy-enhancing technologies, such as secure multi-party computation \cite{goldreich1998secure,keller2020mp}, federated learning \cite{rieke2020future}, and differential privacy \cite{dwork2006differential}, and incorporate them into the framework. Another essential avenue for future investigation lies in understanding the impact of various privacy-preserving mechanisms on the fairness of algorithmic decision-making. Researchers can undertake a comprehensive examination of the tradeoff between privacy and fairness, critically analyzing how privacy-enhancing measures may influence the accuracy, reliability, and equity of algorithmic outcomes.

\section{Conclusion}
\label{sec:discussion}

In conclusion, this work introduces a novel framework for designing fair ubiquitous computing systems, addressing the existing gap in the ubiquitous computing literature on fairness challenges. By presenting this framework, we contribute to the advancement of ubiquitous computing research, with the hope of providing a valuable resource for researchers and practitioners striving to develop more equitable and inclusive systems in the future. We envision that the proposed framework can serve as an initial stepping stone towards fostering fairness and ensuring that ubiquitous computing systems align with ethical principles and societal values. 

\section*{Acknowledgement}
    This material is based upon work supported by the National Science Foundation under Grant No. EDA-2009977 and the University of Washington College of Engineering, Department of Electrical and Computer Engineering, and the Paul G. Allen School of Computer Science and Engineering.

\bibliographystyle{abbrv}
\bibliography{fairness}

\end{document}